# The imaging X-ray detector for Lobster-ISS

J.K. Black [a*], A.N. Brunton [d], N.P. Bannister [d], P. Deines-Jones [b], K. Jahoda [c]

[a] Forbin Scientific, Code 662, NASA/Goddard Space Flight Center, Greenbelt, MD 20771 USA

[b] Universities Space Research Association, Code 661, NASA/Goddard Space Flight Center, Greenbelt, MD 20771 USA

[c] Laboratory for High-Energy Astrophysics, Code 662, NASA/Goddard Space Flight Center, Greenbelt, MD 20771 USA.

[d] Space Research Centre, Department of Physics and Astronomy, University of Leicester, Leicester LE1 7RH, UK.

**Abstract**

Lobster-ISS is a revolutionary astrophysical X-ray all-sky monitor scheduled for deployment as an attached payload on the International Space Station (ISS) in 2009. Using a new wide field-of-view focusing optic, Lobster-ISS provides an order-of-magnitude improvement in the sensitivity of all-sky monitors in the soft X-ray band (0.1 – 3.0 keV). This *lobster-eye* optic presents unique challenges to the imaging X-ray detector at its focus. Micro-patterned imaging proportional counters, with their mechanical simplicity and high performance are the most practical means of meeting the requirements. We describe our design for the Lobster-ISS X-ray imaging detector based on direct-imaging micro-well proportional counters and the expected performance.

## 1. The Lobster-ISS detector requirements

The Lobster-ISS instrument consists of a set of six identical X-ray telescopes mounted on the zenith-pointing external platform of the ISS's Columbus module. Each telescope covers a 27 x 22.5 square degree field-of-view. The instrument's total instantaneous field-of-view is therefore 162 x 22.5 square degrees. With the combination of the wide field-of-view in one direction and the motion of the ISS in the other, Lobster-ISS maps almost the entire sky every 90 minutes. With an angular resolution of four arc minutes, this will generate a confusion-limited catalogue of about 250,000 sources every two months.

Lobster-ISS achieves this remarkable sensitivity by using a wide-field X-ray optic similar to the eye structure of certain crustaceans. The principle of the optic[1] is illustrated in Figure 1. The optic consists of a spherical, square-packed array of channels of square cross section, whose long axes intersect a distance *R* along the axis of symmetry. X-rays entering a channel within the array are brought to a true central focus a distance *R/2* along the symmetry axis, provided they undergo odd numbers of grazing incidence reflections from orthogonal channel walls.

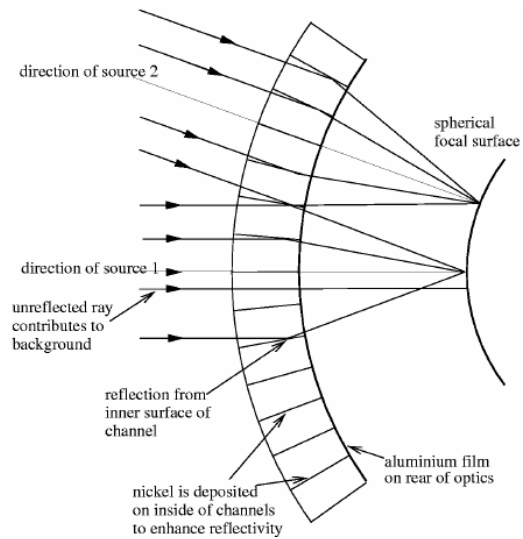

Fig. 1. X-ray focusing using a square pore, spherically slumped microchannel plate. In this case, two sources illuminate the optic.

---

[*] Corresponding author. Tel.: 1-301-286-1231; fax: 1-301-286-1684; e-mail: black@forbinsci.com.

The overall point-spread function is cruciform in nature. Photons reflected an odd number of times in one plane and an even number in the other give rise to line foci which cross at the true focus. X-rays reflected an even number of times in both planes are unfocused and form a diffuse image component. This geometry has been realized in practice with square-pore micro-channel plates (MCPs) at the University of Leicester[2].

The challenge for the X-ray detector at the focus is then clear: covering a large, spherical focal surface with high-resolution imaging. In Lobster-ISS, each of the six telescopes has a 75 cm radius of curvature with a 40 x 40 cm$^2$ area. The focal surface then has a radius of 37.5 cm with an area of 20 x 20 cm$^2$. To achieve the desired 4 arc minute resolution, the total focal spot must be 0.4 mm or less (FWHM). With an expected 3 arc minute spot from the optics, the detector must have image resolution better than about 250 microns (FWHM) over the 0.1 – 3keV bandpass.

## 2. Lobster-ISS detector design

The rapidly maturing technology of micro-pattern gas detectors[3] provides a practical means to achieve the X-ray imaging performance required by Lobster-ISS. Combined with modern analog application-specific integrated circuits (ASICs), micro-pattern detectors provide high-resolution imaging in the soft X-ray band over large areas with low power consumption and modest cost. Micro-pattern detectors fabricated on polymeric substrates also have the mechanical simplicity and robustness needed to approximate the spherical focal surface sufficient to meet the spatial resolution requirements.

Each of the six Lobster-ISS telescopes has at its focal surface four 10 x 10 cm$^2$ imaging micro-pattern detectors with sixteen flat windows of approximately 25 cm$^2$ each. A spherical detector is approximated by arranging the planar detector panels and windows in the form of a geodesic with the focal surface radius.

### 2.1. Imaging micro-well detectors

The micro-pattern detectors are imaging micro-well detectors. The basic detector geometry, shown in Figure 2, is like that of the micro-CAT[4,5] or WELL[6]. The detector is an array of charge-amplifying cells placed below a drift electrode, which also usually serves as a detector window. The sensitive volume is a gas-filled region between a drift electrode above the cathode and the micro-well.

Each cell of a micro-well detector consists of a charge-amplifying micro-well. The well itself is a cylindrical cavity formed in a polymeric substrate. The cathode is a metal annulus around the opening of the well while the anode is a metal pad covering the bottom of the well.

Primary ionization electrons created by X-ray interaction in the sensitive volume drift toward the micro-well and are swept into the wells toward the anode. In the well, the electric fields are strong enough to create an ionization avalanche. Electrons from the avalanche are collected at the anode, while an equal, but opposite, charge appears on the cathode.

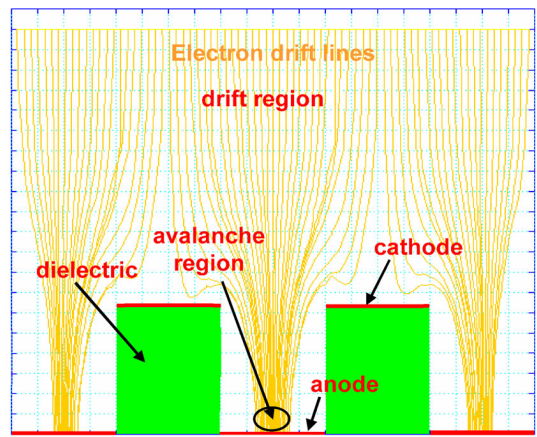

Fig. 2. Cross section of an array of micro-wells. The anodes efficiently collect the ionization electrons. Equal but opposite charges are generated on the anodes and cathodes.

Connecting the cathodes and anodes in rows and columns forms an orthogonal coordinate system, so that the micro-well detector becomes an imaging system. Reading out each row and column then acquires an image directly.

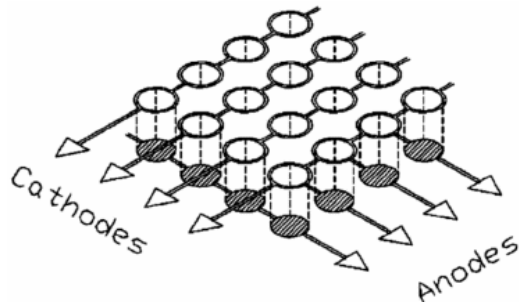

Fig. 3. Micro-well anodes and cathodes read out in rows and columns form an orthogonal coordinate system for imaging.

The baseline detectors for Lobster-ISS are micro-wells with rows and columns on a 400 micron spacing, made on a 125 micron polyimide substrate. Such detectors have demonstrated stable operation at gas gains up to $3 \times 10^4$, 20% energy resolution at 6 keV, and 85 micron ($\sigma$) position resolution[7].

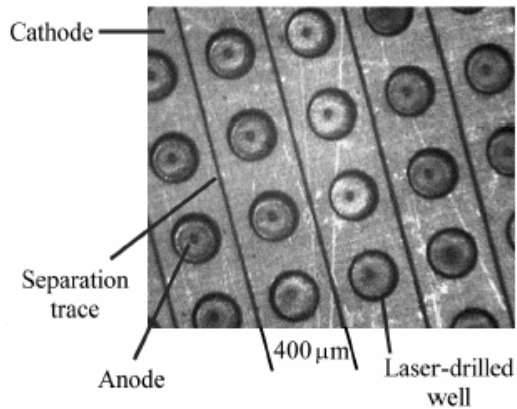

Fig. 4. Optical microscope image of the cathode face of a micro-well detector fabricated with UV-laser micro-machining on a 125 micron thick copper-coated polyimide substrate.

### 2.2. Readout electronics

The detector readout electronics are based on a system developed for double-sided silicon strip detectors at the Max-Planck-Institut fur Extraterrestrische Physik (MPE)[8].

The system is based on a commercial ASIC (the TA1) that is a self-triggered, 128-channel charge-sensitive preamplifier-shaper with simultaneous sample-and-hold and a multiplexed analog output. The TA1 has a shaping time of about two µsec and a minimum readout time of 12.8 µsec. The MPE system also refers the ASIC ground to the electrode bias potential, thus eliminating the need for high-voltage coupling capacitors. Analogue data and control signals are transmitted through optocouplers.

With a power consumption of about 0.5 mW/channel, the ASIC has a dynamic range of about $1.4 \times 10^5$ electrons and can be configured to accept signals of either polarity. At a shaping time of 2 microseconds, the rms baseline noise is 165 electrons with a noise slope of 7 electrons/pF. With a micro-well strip capacitance of about 6 pf, the rms noise is such that the detectors can be operated at a modest gain of $\sim 10^3$, far below the maximum stable gain of $\sim 3 \times 10^4$, and remain linear up to the maximum of the Lobster bandpass (3.0 keV).

Each $10 \times 10$ cm$^2$ detector then has four ASICs, two for anodes and two for cathodes. There are then 96 ASICs for the entire instrument with a total power consumption of 6W for the front-end electronics.

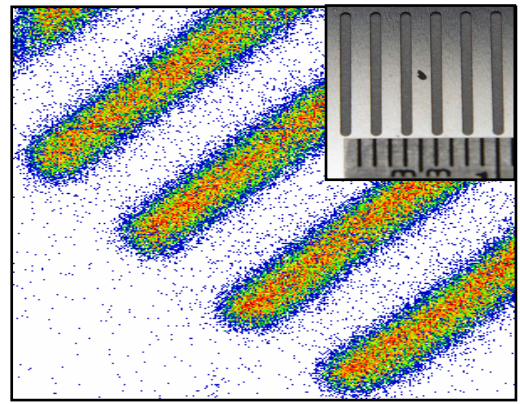

Fig. 5. False-color image of a shadow mask (inset) illuminated with 3 keV X-rays made with a 400-micron pitch micro-well. The mask has 750 micron-wide slots on a 2mm pitch.

*2.3. Thin detector window*

Large-area thin windows are an important aspect of the Lobster-ISS detectors. The baseline detector windows are one-micron thick polyimide films, coated with Al/AlN. The estimated leak rate for such windows is ~7 x $10^{-6}$ $cm^3$ $cm^{-2}$ $s^{-1}$, and an auxiliary gas reservoir will supply sufficient gas for a 3-year period of operation. We are investigating 0.2 micron thick silicon nitride films as windows. These would eliminate the need for a gas replenishment system and provide 40% higher transmission below 1.8 keV.

## 3. Expected imaging performance

In addition to the intrinsic position resolution of the finite-pitch micro-well detector, there are components of the position error arising from diffusion, the finite thickness of the detector, and the fact that the flat detectors only approximate the shape of the focal surface.

Diffusion produces an error in the measured centroid of the charge cloud. This contribution is most significant for low-energy X-rays, and is limited by reducing the detector thickness, at some cost to detector efficiency at high energies.

Focusing error occurs whenever X-rays are not absorbed at the focal surface. De-focusing has an energy-independent component because planar detectors cannot everywhere conform to the focal sphere, and an energy-dependent term because absorption depth increases with energy.

Parallax error, as we define it, occurs because the drift field from planar electrodes is not everywhere radial. This results in rays from the optic cutting across drift cells.

The total estimated detector error from the baseline detector – 10 cm windows, 3 mm thickness, 1.1 Bar Xe – is plotted in Figure 6. The total resolution is dominated everywhere by the intrinsic resolution of the detector. Based on these estimates, the baseline detector meets or exceeds the resolution requirement of 250 microns (FWHM). Further improvements in resolution, if desirable, would be most easily gained by reducing the detector pitch.

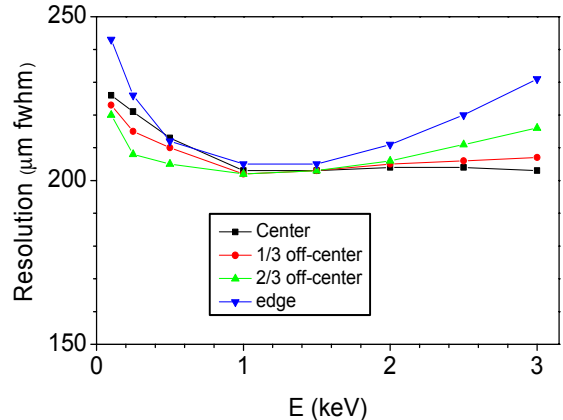

Figure 6. Estimated position resolution for the baseline Lobster-ISS detectors plotted for various locations across the face of a detector. The resolution exceeds the 250 micron requirement.

**Acknowledgments**

The authors would like to thank G. Kanbach, F. Schopper, and R. Andritschke and the Max-Planck-Institut fur Extraterrestrische Physik for their very generous assistance developing a readout system suitable for Lobster-ISS. This work was supported by the NASA's Goddard Space Flight Center.